# Fundamentals of ion mobility spectrometry

Valérie Gabelica,[1] Erik Marklund[2]

1. Univ. Bordeaux, INSERM, CNRS, Laboratoire Acides Nucléiques Régulations Naturelle et Artificielle (ARNA, U1212, UMR5320), IECB, 2 rue Robert Escarpit, 33607 Pessac, France. E-mail: v.gabelica@iecb.u-bordeaux.fr

2. Department of Chemistry – BMC, Uppsala University, Box 576, 75123, Uppsala, Sweden. E-mail: erik.marklund@kemi.uu.se

## Abstract

Fundamental questions in ion mobility spectrometry have practical implications for analytical applications in general, and omics in particular, in three respects. (1) Understanding how ion mobility and collision cross section values depend on the collision gas, on the electric field and on temperature is crucial to ascertain their transferability across instrumental platforms. (2) Predicting collision cross section values for new analytes is necessary to exploit the full potential of ion mobility in discovery workflows. (3) Finally, understanding the fate of ion structures in the gas phase is essential to infer meaningful information on solution structures based on gas-phase ion mobility measurements. We will review here the most recent advances in ion mobility fundamentals, relevant to these three aspects.

## Highlights

- CCS distributions and shifts contain valuable information about analytes
- Compound-dependent effects of gas, field and temperature explain calibration issues
- New algorithms improve accuracy and speed of collision cross section calculations
- Gas-phase charge sites or conformations do not always reflect the solution ones

# Introduction

Omics sciences require to separate, identify and quantify all compounds in a mixture. Ion mobility spectrometry (IMS), in which electric fields are used to drag analytes through a buffer gas, is useful for separation sciences, and can also aid identification. We review here the fundamental principles behind using IMS for identification and structural characterization. Although theoretical foundations were laid decades ago [1,2], fundamental contributions have flourished in the last two years. We gather contributions of prime importance for IMS, spanning from small molecules (lipidomics, metabolomics) to large multi-protein assemblies (structural proteomics and native mass spectrometry).

In IMS, the force exerted by an electric field on an analyte ion is exactly balanced by friction with the buffer gas, yielding a steady-state analyte velocity $v_d$. The ion mobility $K$ (Eq. 1) is thus a measure of friction linked to an observable, the time $t_d$ the ions take to traverse the length $l$ of the mobility cell.

$$K = \frac{v_d}{E} = \frac{l}{t_d E} \tag{1}$$

$K$ depends on the collision frequency, hence on the gas number density ($N$), gas temperature ($T$) and pressure ($p$), so the reduced mobility $K_0 = K.N/N_0 = K.(p/p_0).(T_0/T)$ is better to compare different experiments (in standard conditions, $N_0 = 2.687 \times 10^{25}$ m$^{-3}$, $p_0 = 760$ Torr, $T_0 = 273.16$ K). When $v_d$ is small compared to the ion thermal velocity $v_T$, $K$ can be expressed as Eq. (2) [1].

$$K = \frac{3}{16}\sqrt{\frac{2\pi}{\mu k_B T}} \frac{ze}{N\Omega} \tag{2}$$

$\mu$ is the reduced mass of the ion–gas pair ($\mu = mM/(m+M)$, where $m$ and M are the ion and gas-particle masses), $k_B$ is the Boltzmann constant, and $ze$ is the analyte charge.

$\Omega$, often called the "collision cross section" (CCS), is actually a "momentum transfer collision integral", that is, the momentum transfer between ion and gas particles averaged over all gas-ion relative thermal velocities. While the terms tend to be used interchangeably in IMS, they are in fact not identical in a wider context. Scattering or dephasing measurements carried out at very low pressures, wherein collisions eject the ions from stable trajectories [3-6], allow to determine true scattering collision cross sections, which can be adequately calculated by a projection approximation. Ion mobility is different: we still detect the ions after they had undergone collisions. The momentum transfer collision integrals measured in ion mobility are different, and require taking into account the effects of the gas on the ion momentum (i.e., velocity). Although CCSs and momentum transfer collision integrals are related, they are thus not necessarily identical, and further work is warranted to bridge the gap between the two types of experiments.

In ion mobility (Eq. 2), $\Omega$ has the dimensions of a surface, is a property of the ion–gas pair, and also depends on other parameters influencing the ion–gas collision velocities, i.e. on the temperature $T$, on the electric field $E$ and on the pressure $p$ (which controls $N$), and specifically on $E/N$. The first section will review the effects of gas, field and temperature, which are crucial to interpret the data correctly, and to understand differences between instrumental setups.

IMS practitioners have three main ways to characterize the analytes: $t_d$ (in practice, the arrival time at a detector, $t_A$), $K_0$, and $\Omega$. Each of these values can aid identification of "*known knowns*" (molecules anticipated by the researcher, by comparison with a measured database) or of "*known unknowns*" (compounds that are unknown to the researcher, but described in the literature, by comparison with a predicted database). As $t_d$ or $t_A$ values depend on the instrument and on experimental conditions, they have only in-house utility. In contrast, databases of mobilities or cross section values are in principle transferrable, and the conditions for their transferability across instrumental platforms will be discussed below. Moreover, $\Omega$ can aid the identification of "*unknown unknowns*", by comparing values predicted from putative candidate structures. The fundamentals of CCS calculation will be covered in the second section. Finally, we highlight recent examples of how IMS measurements and modeling shed new light on one of the most fundamental questions of mass spectrometry: how the structure in the gas phase relates to those in solution.

**Effect of drift gas on collision cross sections**

Early IMS for structural elucidation was carried out in drift tubes (DT), operated in helium because calculations are easier. Using IMS for omics became possible with the introduction of commercial high-performance electrospray IMS mass spectrometers, usually operated in nitrogen. The first commercial IMS (the Synapt HDMS™, introduced by Waters in 2006) operates with traveling wave (TW) IMS [7]. Because the electric field is not static in TWIMS, apparent drift times have not the same meaning as drift tube $t_d$ values. An empirical correlation was made to match arrival times based on helium drift-tube CCSs [8], and recently the $t_d$ of peptides and proteins could be modeled at low wave velocities directly from $K$ without calibration [9].

Experience, however, showed that TWIMS calibration is not universal: CCS values depend on the nature of the calibrant, which should have size, charge, and chemical class similar to the analytes (doing so results in an average deviation of < 2% between $^{TW}\Omega_{N2 \to He}$ and $^{DT}\Omega_{He}$ [10]). Similarly, native protein complexes should be used for structural proteomics [11]. However, Konermann's group recommends using denatured proteins even for native protein analytes, because denatured proteins do not change CCS values when changing pre-IMS activation conditions [12]. Recently, the Robinson group also showed that even native soluble proteins are inappropriate calibrants for native membrane proteins [13]. A hot question for TWIMS users is thus: "*What makes a calibrant suitable for my analytes?*"

Helium drift tube CCS values are often used to calibrate TWIMS instruments operated in nitrogen, so let's first discuss the effects of the drift gas on the CCS. Benzocaine will serve as textbook example for small molecules. In positive-mode electrospray, benzocaine forms two tautomers (different proton location, but same conformation) [14••]. They are readily separated by IMS in nitrogen, but overlap in helium (Figure 1). At 300K, interactions between the ions and the helium are akin to collisions with a hard sphere, hence the mobility difference is small. In contrast, nitrogen is polarizable, and interacts more strongly with the more polar tautomer [15]. The proportionality factor between helium and nitrogen CCS values thus depends on the chemical nature (here, charge location) of the analyte, and this effect is strongest for small ions.

Recent simulations of how $\Omega$ depends on the gas temperature and identity highlight how the trends depend on the ion charge [16] and shape [17••] (Figure 2). $\Omega$ increases when the temperature decreases, because long-range interactions become more dominant at reduced $v_T$ [18,19]. At high temperature, the impulses from "grazing" gas particle collisions are smaller in magnitude as high thermal velocity allows little time for the interatomic forces to act, and $\Omega$ approaches the hard-sphere limit. The $\Omega$-difference between gases at high temperature is due to the gas-particle radius, and correlates with ion size and shape [17••]. However at lower temperatures, *including room temperature*, the difference between He and $N_2$ is further influenced by long-range interactions, which depend on the gas polarizability, ion net charge [16] (Figure 2, top), and exact charge localization as illustrated by the benzocaine example [14••] (for proteins, one report claims that this effect seems negligible [20], an another claims that it is significant in helium for high charge states [21]). Finally, the ion shape also matters (Figure 2, bottom). In the case of proteins, denatured conformations expose more residues than globular ones, and this increases the interactions with the gas.

In summary, the ratio between $\Omega_{N2}$ and $\Omega_{He}$ depends on (i) the temperature, (ii) the ion charge, (iii) the charge localization, (iv) ion size and (v) ion shape. Yet, helium-to-nitrogen CCS conversion does not suffice to explain all compound-class effects on empirical TWIMS calibrations: even when calibrating TWIMS with drift tube CCS values measured in nitrogen, compound class still matters [13,22]. One reason can be that second-generation TWIMS instrument do not operate in pure nitrogen either, but in a nitrogen/helium mixture (helium coming from the pre-IMS cell). Another explanation lies in differences in field heating regimes, combined with the temperature dependence of nitrogen CCSs.

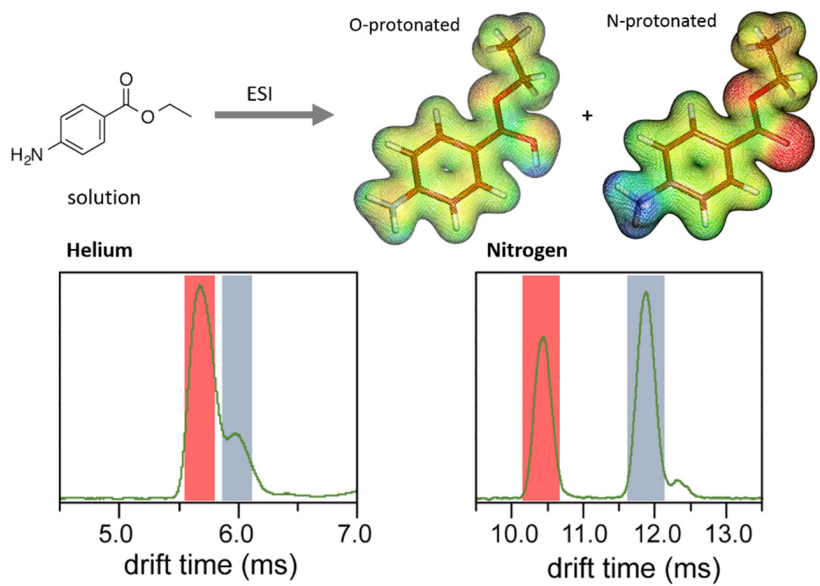

**Figure 1:** The O-protonated and N-protonated forms of benzocaine, produced simultaneously by electrospray in acetonitrile, separate differently in helium or nitrogen drift tube ion mobility (image courtesy of Kevin Pagel [14••]).

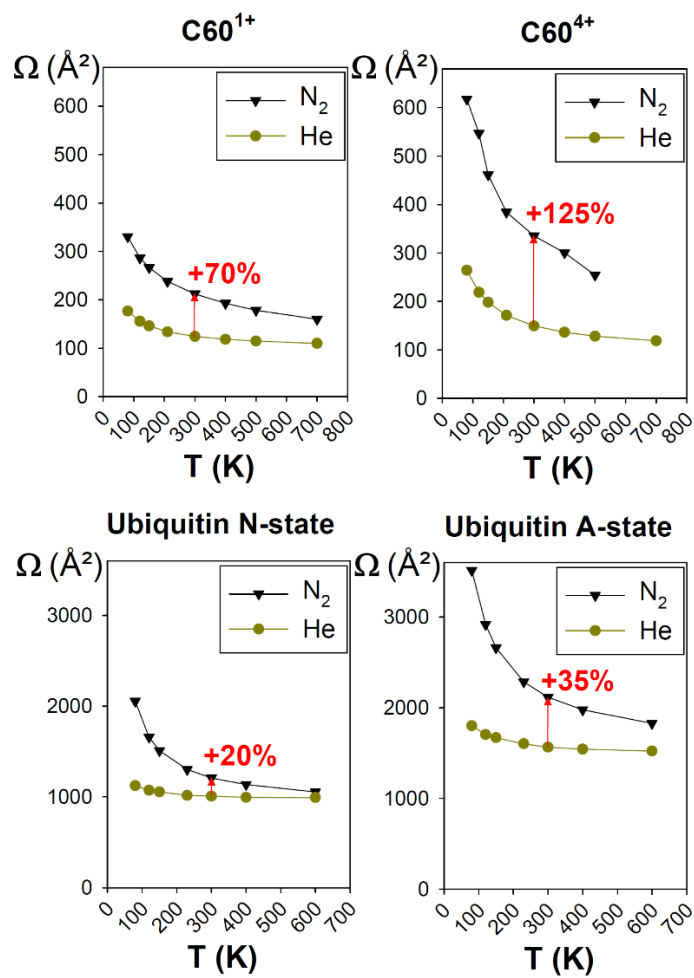

**Figure 2:** Calculated temperature dependence of CCSs in $N_2$ and He (top) for $C_{60}$ at two different charge states, using the trajectory model [16], and (bottom) for ubiquitin in two different conformations (N-state is native, A-stated is less compact), using the projection superposition approximation [17$^{\bullet\bullet}$].

## Are we measuring ion mobility at the low-field limit?

Understanding the $E/N$ effect is crucial to determine whether $K_0$ or $\Omega$ values can be compared across different platforms, the confidence with which databases can be shared, and the expected error or bias. Equations (1-2) are usually presented without questioning their validity. However, the reduced ion mobility $K_0$ is not a constant, because it depends on the $E/N$ ratio (expressed in Townsends; 1 Td = $10^{-21}$ V.m²) [23]:

$$K_0\left(\frac{E}{N}\right) = K_0(0)\left[1 + \alpha_2\left(\frac{E}{N}\right)^2 + \alpha_4\left(\frac{E}{N}\right)^4 + \cdots\right] \tag{3}$$

The low-field limit means that $E/N$ is small enough so that $K_0$ is independent of $E/N$. High-field asymmetric waveform ion mobility spectrometry (FAIMS) exploits this dependence of mobility on $E/N$ to separate ions, and important discussions on the onset of the high-field limit can be found in the FAIMS literature [24]. These discussions were left aside in most of the drift tube and traveling wave IMS literature, but are now resurfacing because, as the precision and accuracy of IMS is improving, significant effects can be detected at lower fields than thought previously [25,26].

Importantly for all those using collision cross section values, Eq. (2) is valid only in the low-field limit. The value of the maximum electric field at which the instrument can operate without affecting the ion mobility in a noticeable way depends on the ion-gas pair. For atomic ions in noble gases, the low-field limit is of the order of 10 Td [24] but recently, $E/N$ effects for polyatomic ions were reported below 4 Td [25,26]. Such ongoing research matters, because typical $E/N$ values of modern high-performance IMS instruments may well be outside of the low-field limit (Table 1).

A key recent contribution is the momentum transfer theory [27,28]**, which improves Equation (2) for non-zero fields and for ions significantly heavier than the buffer gas (thus, adapted to omics applications). Equation (2) is valid only at electric fields weak enough so that the drift velocity $v_D$ is small compared to the thermal velocity $v_T$ at zero field (Eq. 4).

$$v_T = \sqrt{\frac{8k_B T}{\pi \mu}} \tag{4}$$

For $T$ = 300 K and $m$ = 200, $v_T$ is 1239 m/s in helium and 495 m/s in dinitrogen. In Table 1, we see that in traveling wave IMS, $v_D$ becomes dangerously close to $v_T$, and thus Equation (2) is not necessarily valid. The magnitude of the ensuing errors remains to be determined.

Moreover, working above the low-field limit means that field heating could change the effective temperature of the ions, resulting in measurable fragmentation [29,30] or isomerization [31] as shown for ions of 200-300 Da in TWIMS. In the previous section, we saw that (1) nitrogen CCSs vary with the temperature in the 300-500K range, and that (2) the analyte chemical class influences the *T*-dependency. Class-dependent calibration problems in TWIMS could thus also come from ion temperatures in TWIMS differing from the calibrant DTIMS temperatures. In summary, DT and TW experiments will transpose well only if analytes and calibrant CCS values have the same temperature dependency in the buffer gases of interest. This condition is more likely to be met if calibrant and analyte are similar in charge, charge localization (depending on analyte size), size and shape. We hope further studies, in particular with temperature-dependent drift tubes [32], will help test our proposal.

**Table 1. Estimates of typical operating parameters for contemporary commercial instruments.**

| Instrument and manufacturer | Operation principle | *p* (Torr) | *E* (V/cm) | Typical *E/N* (Td) | Typical $v_D$ (m/s) |
|---|---|---|---|---|---|
| 6560 IMS-Q-TOF, Agilent Technologies | Drift tube | 4 | 9.5-20 | 7-15 [33] | 10-80 |
| Synapt HDMS, Waters | TWIMS | 0.4 | 21 (maximum axial field at wave height = 10 V) [34] | ≤160 | 200-600 [29] |
| Synapt G2, G2-S and G2Si HDMS, Waters | TWIMS | 2 | 100 (maximum axial field at wave height = 40 V) | ≤155 | 200-300 [30] |
| TIMS-TOF, Bruker | TIMS | 2 | 30-55 | 45-85 | 120-170* [35] |
| IMS-TOF, Tofwerk | Drift tube | 570-788 | ~400 | 1-2 [36] | ~5 |

\* Gas velocity; in TIMS the ion is static.

## Structural interpretation based on collision cross sections

Since the information contained in the CCS alone is insufficient to uniquely define the analyte structure, molecular modelling plays an important role in structural interpretation of IMS data, i.e., for unknown unknowns. The approach, conceived almost a century ago [37], is to calculate CCS values for candidate structure models and compare them with experimentally derived CCS values. Several methods have emerged (Table 2 and Figure 3). Most of them shoot buffer-gas probes toward a (static) analyte in a Monte-Carlo integration of the collision integral. Their main differences stem from what trade-offs between speed and physical rigor follow from their underlying assumptions.

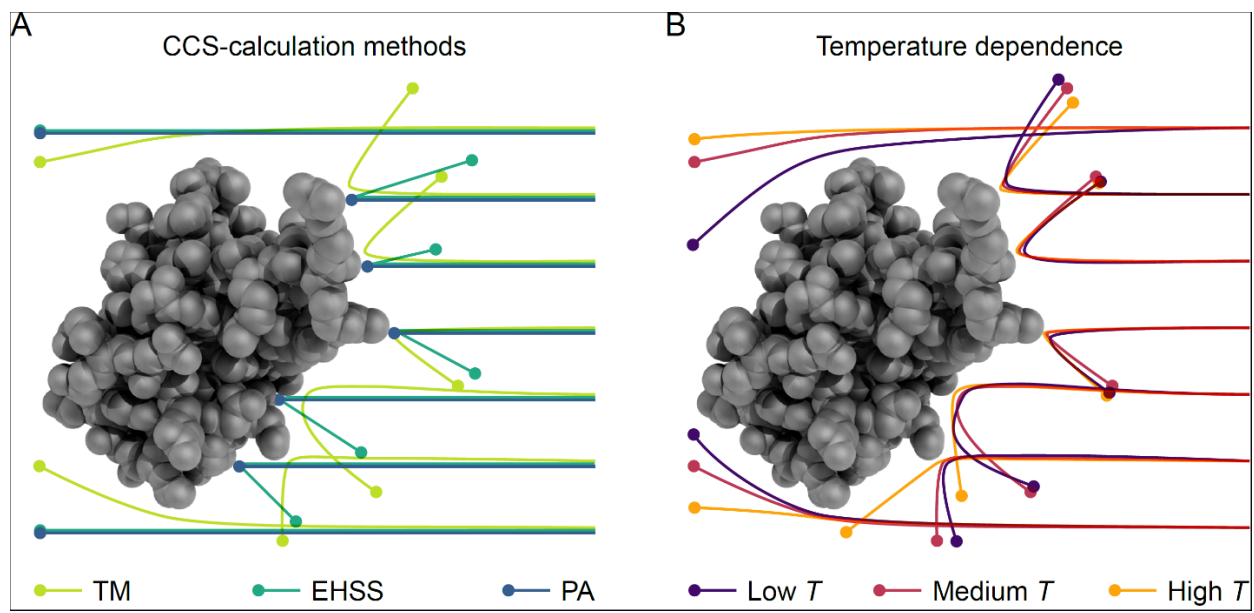

**Figure 3:** A) The principal operation and differences between the main classes of CCS calculation methods. TM traces the probe particles' trajectories, whereas EHSS and PA only regards direct contact between probes and analyte. The EHSS considers (multiple) scattering after initial collision, whereas PA only differentiates between hits and misses, and estimates CCSs from the fraction of hits. B) At higher temperatures, the momentum transfer between buffer gas and analyte effectively decreases. This is easily seen for 'grazing' collisions in the figure, which are deflected more as they pass the analyte.

The Trajectory Method (TM) treats long- and short-range interactions (we saw above that this is necessary for small molecules in nitrogen) and the $v_T$ distribution explicitly [19]. The numerous force evaluations for each trajectory, combined with the vast integration domain, makes TM extremely time consuming (Table 2) [38••,39,40]. A solution is to replace trajectory calculations by a local collision probability approximation (LCPA) [41]. Adaptation for molecular buffer gasses [42-45] strengthens TM's fundamental position for CCS calculation. TM in its canonical form assumes elastic collisions, which is not a fully valid assumption for molecular buffer gases, and extension to accommodate inelastic scattering is an important enhancement [46].

For larger analytes, long-range interactions become less influential on average [16,42], allowing for further approximations. The Elastic Hard Sphere Scattering (EHSS) method [47] disregards long-range interactions, considering only hard-sphere collisions. An improved way to define the ion surface was presented in the Scattering on Electron Density Isosurfaces (SEDI) method [48,49]. The Projection Approximation (PA) completely ignores gas scattering effects, assuming the CCS to be the analyte's rotationally averaged projected area (taking the buffer-gas particle radius into account) [37], and for large analytes such as proteins the PA CCS is thus an (inverse) metric of the analyte's self-occlusion [50]. These methods require no integration of individual probe trajectories and are considerably faster than TM [38••,39,40], yet can surprisingly give results within a few percent of TM values [20,38••,39], either through calibration [38••] or by their own accord if the effects of surface roughness average out. Domain-decomposition schemes give PA a performance boost for large structures [38••,51], and future gains might be made by exploiting mathematical properties of projected areas [50]. Similar enhancements are conceivable also for the more rigorous methods, but remain to be implemented. Omission of long-range interactions make EHSS and PA invariant to temperature, although temperature-dependent atom radii can circumvent this problem to some degree [39,52]. PA lends itself for analysis of non-atomistic structures, including bead models [53] and electron-densities [38,54]. Building on PA, Projected Superposition Approximation (PSA) employs a mean-field approximation for scattering, based on the surface characteristics of the analyte, and approximates long-range interactions via distance- and temperature-dependent collision probability [39]. PSA is more able than PA to handle non-globular structures, at the cost of four orders of magnitude in throughput (Table 2).

Machine learning using various molecular descriptors offers a radically different alternative to calculate the (approximated) collision integral. For small molecules this yields accurate CCSs for approximately 90% of tested analytes [55••,56••], outperforming TM [56••]. Some descriptors are computationally expensive, but less so than TM calculations [55••], and omitting geometric properties drastically improves throughput [56••]. This approach is not yet available for protein-sized analytes, but could be a powerful alternative if such parameterization is possible.

**Table 2. Methods for calculating CCSs of a given structure, their performance, and available implementations. Performance is given as the approximate time for computing the CCS of a protein structure to a statistical error of 1%.**

| Method | Computing time (s) | Available implementations |
| --- | --- | --- |
| TM | 63 100 [38••] | Mobcal [19], ImOS* [46], Collidoscope [57] |
| EHSS | 1670 [38••] | Mobcal [47], EHSSrot [58], ImOS* [46] |
| PA | 0.07 [38••] | Mobcal [19,47], ImOS [46], Sigma† [59], Impact [38••], CCSCalc [60] |
| PSA | 663 [39] | - |

*) ImOS additionally implements diffuse scattering versions of TM and EHSS – DTM and DHSS – which extends the physical model to take non-elastic collisions into account. †) Implements a mass-dependent scaling factor to take into account the collective long-range interactions from many atoms.

## Do gas-phase ion structures reflect the solution phase ones?

The gas phase structure can differ from the solution ones, either because of the ionization, or because of a different balance of interactions. In physical chemistry terms, this is represented by different potential energy surfaces (PES) for the solution and the gas phase [61]. With low internal energy (i.e. just enough for desolvation and declustering) one hopes to trap conformations that have kept some memory of the solution ones [61]. Large proteins and protein complexes are likely to preserve salt bridges [62], but smaller analytes can retain structural elements as well. For example, IMS was successfully used to study the solution biophysics of cis-trans isomerization of polyproline [63,64•]. Even small pentapeptides can adopt kinetically trapped conformations in the gas phase [65], so the internal energy dependence of gas-phase conformations can be relevant to proteomics workflows as well. The experiment time scales are a determining factor for the survival of native-like states. The typical IMS experiment take micro- to milliseconds, with little interconversion between distinguishable protein conformers [66], whereas for longer times the fraction of extended conformations increases [67].

Still, in our opinion, representing a single gas-phase PES is still an oversimplification, because there are as many co-existing potential energy surfaces as there are ways to locate the charges. We have seen that for small molecules, a single analyte in solution can give two mobility peaks corresponding to two possible protonation sites in the gas phase. This is an important fundamental insight for metabolomics interpretation, and a warning that the number of "features" observed in IM-MS is not necessarily equal to the number of distinct compounds in the sample. For larger, multiply charged ions, it would be further useful to distinguish charge macrostates (total charge) from charge microstates (a particular distribution of charge locations). In addition, molecular flexibility favors proton migrations in the gas phase, also at room temperature, as shown for protein polycations [68•], glycan monoanions [69] and oligonucleotide polyanions [70]. In turn, proton migration allows the ion to explore different microstates with other conformations. The ion mobility community could be inspired by efforts made to include charge migration in molecular modelling of gas-phase structural ensembles in the framework of understanding collision-induced asymmetric charge partitioning of multiprotein complexes [71-74]. Further complicating the assignment based on structural modeling, fast interconverting structures have drift times corresponding to their abundance-weighted average CCSs, not to CCSs of individual conformations [69,75]. Cations ($Na^+$, $Ca^{2+}$,…) are less mobile at room temperature, thereby freezing conformations. This helps for some analytical applications, for example in carbohydrate analysis [76].

As the analytes become larger (e.g., proteins), the width of the charge-state and CCS distributions are generally correlated with flexibility and disorder [77-79]. Broader-than-diffusion ion mobility peaks result from flexible structures rearranging in a myriad different ways and coexisting without interconverting after transfer to the gas phase, to self-solvate their charges [80•]. Broad peaks obtained from native proteins [66] indicate that this phenomenon is not restricted to intrinsically disordered structures and warrants attention. This is purely a gas phase process, as demonstrated by IMS on low-charge state proteins produced by neutralization of high charge-state ones [81,82]. However, collision-induced unfolding allows to differentiate conformational ensembles from the way they change with internal energy [83]. When the total charge repulsion overcomes the intramolecular binding forces including salt bridges (akin to an intramolecular Rayleigh limit), large flexible molecules elongate to conformations much more extended than in solution [84,85]. For these reasons, intrinsically disordered proteins adopt both more compact and more extended (depending on the charge state) conformations in the gas-phase compared to the solution [86••]. Finally, note that a compact conformation does not necessarily mean that the solution structure is preserved, because at low charge states, collision-induced rearrangements can lead to compaction as well [87].

## Concluding remarks

Over recent years IMS has boosted fundamental studies on ion structures in the gas phase, and their relationship with the electrospray process (charging and mode of ejection from the droplets) and with internal energy. The relationship between solution and gas-phase structure is not always straightforward, but we anticipate that this lively area of research will help us rationalize many puzzling observations and uncover principles that can lead to routine application of ion mobility in omics studies.

Reflecting the complexity and the dependence on experimental detail, we propose to go beyond the reporting of single CCS values. Measurements in multiple buffer gases [17••], at different temperatures, or with different electric fields (for example, using Field Asymmetric Ion Mobility Spectrometry (FAIMS)), once the fundamentals will be understood, will paint a more complete picture of the analyte. The differences in CCS values, instead of being an annoyance for inter-instrument comparisons, would rather provide additional information for identification.

On the modelling side, we would welcome development of force fields better suited for MD under gas-phase conditions, possibly including polarizability, and further development of methods for dynamic proton-transfer; both of which would enable more accurate studies of fundamental aspects of IMS and MS. For CCS calculations, we would like to see further exploration of when the more approximate methods can be used in place of TM, their proper parameterization, and how they can be calibrated. The throughput difference between TM and PA is about a million-fold, with clear implications on the number of structure models that can be tested, which translates to the strength of the interpretations made. Inelastic collisions and the impact of gas-phase dynamics on the CCSs are also pertinent aspects, where CCS calculations and MD simulations can combine.

As this review has outlined, many fundamental questions have been answered over recent years. Many remain however, and the coming years have potential to bring great progress to IMS for omics applications.

**Acknowledgements**

We acknowledge the funding from the European Union's 7th Framework Programme (E.M. for a Marie Skłodowska Curie International Career Grant, 2015-00559, V.G. for an ERC Consolidator grant DNAFOLDIMS, project number 616551) and the Swedish Research Council. We thank the COST Action BM1403 WG4 for fruitful discussions, and Kevin Pagel for providing us with the images for Figure 1.